\documentclass[aps,prb,twocolumn,superscriptaddress,nofootinbib,floatfix,letterpaper]{revtex4-1}

\usepackage{euscript}

\newcommand\eC           {\EuScript{C}}

\newcommand\eM          {\EuScript{M}}

\newcommand\Bulk       {\mathsf{Bulk}}
\newcommand\sC         {\mathsf{C}}
\newcommand\sD         {\mathsf{D}}
\newcommand\sE         {\mathsf{E}}
\newcommand\sM         {\mathsf{M}}
\newcommand\sGT         {\mathsf{GT}}

\newcommand{\Rep}{{\cR\mathrm{ep}}}
\newcommand{\sRep}{{\mathrm{s}\cR\mathrm{ep}}}

\begin{document}

\begin{titlepage}
\title{A systematic construction of gapped non-liquid states
}

\author{Xiao-Gang Wen}
\affiliation{Department of Physics, Massachusetts Institute of
Technology, Cambridge, Massachusetts 02139, USA}

\begin{abstract} 

Gapped non-liquid state (also known as fracton state) is a very special gapped
quantum state of matter that is characterized by a microscopic cellular
structure. Such microscopic cellular structure has a macroscopic effect at
arbitrary long distances and cannot be removed by renormalization group flow,
which makes gapped non-liquid state beyond the description of topological
quantum field theory with a finite number of fields.  Using Abelian and
non-Abelian topological orders in 2-dimensional (2d) space and the different
ways to glue them together via their gapped boundaries, we propose a systematic
way to construct 3d gapped states (and in other dimensions). The resulting
states are called cellular topological states, which include gapped non-liquid
states, as well as gapped liquid states in some special cases.  Some new
fracton states with fractal excitations are constructed even using 2d $\Z_2$
topological order.  More general cellular topological states can be constructed
by connecting 2d domain walls between different 3d topological orders. The
constructed cellular topological states can be viewed as fixed-point states for
a reverse renormalization of gapped non-liquid states.

\end{abstract}


\maketitle

\end{titlepage}

{\small \setcounter{tocdepth}{1} \tableofcontents }

\section{Introduction}

Different phases of matter are not only characterized by their symmetry
breaking patterns.\cite{L3726,L3745} Even systems without any symmetry can have
many distinct gapped zero-temperature phases, characterized by different
patterns of long range quantum entanglement\cite{CGW1038}.  Those gapped phases
include \emph{gapped liquid phases}\cite{ZW1490,SM1403} [which incluse phases
with topological orders\cite{W8987,W9039,KW9327}, symmetry enriched topological
(SET)
orders\cite{W0213,EH1306,HW1351,X13078131,MR1315,CF14036491,CY14126589,CQ160608482,HL160607816}
and symmetry protected trivial (SPT) orders\cite{GW0931,CLW1141,CGL1314}], as
well as \emph{gapped non-liquid phases}\cite{C0502,H11011962}, such as foliated
phases\cite{ZW1490,SC171205892,SC180310426} (\ie type-I fracton
phases\cite{VF160304442}).

So far, we have a nearly complete understanding \emph{gapped liquid phases} for
boson and fermion systems with and without symmetry.  In 1+1D, all gapped
phases are liquid phases. They are classified by
$(G_H,G_\Psi,\om_2)$\cite{CGW1107,SPC1139}, where $G_H$ is the symmetry group
of the Hamiltonian, $G_\Psi$ the symmetry group of the ground state $G_\Psi
\subset G_H$, and $\om_2\in H^2(G_\Psi,\R/\Z)$ is a group 2-cocycle for the
unbroken symmetry group $G_\Psi$.  

In 2+1D, we believe that all gapped phases are liquid phases. They are
classified (up to $E_8$ invertible topological orders and for a finite unitary
on-site symmetry $G_\Psi$) by $(G_H,\Rep(G_\Psi) \subset\eC \subset\eM)$ for
bosonic systems and by $(G_H,\sRep(G_\Psi) \subset\eC \subset\eM)$ for
fermionic systems\cite{BBC1440,LW160205946,LW160205936}.  Here $\Rep(G_\Psi)$
is symmetric fusion category formed by representations of $G_\Psi$, and
$\sRep(G_\Psi)$ is symmetric fusion category formed by $\Z_2^f$-graded (\ie
fermion graded) representations of $G_\Psi$.  Also $\eC$ is a braided fusion
category and $\eM$ is a minimal modular
extension\cite{LW160205946,LW160205936}.

In 3+1D, some gapped phases are \emph{liquid phases} while others are
\emph{non-liquid phases}.  The 3+1D gapped liquid phases without symmetry for
bosonic systems (\ie 3+1D bosonic topological orders) are classified by
Dijkgraak-Witten theories if the point-like excitations are all bosons, by
twisted 2-gauge theory with gauge 2-group $\cB(G,\Z_2)$ if some point-like
excitations are fermions and there are no Majorana zero modes, and by a special
class of fusion 2-categories if some point-like excitations are fermions and
there are Majorana zero modes at some triple-string
intersections\cite{LW170404221,LW180108530,ZW180809394}.  Comparing with
classifications of 3+1D SPT orders for bosonic\cite{CGL1314,K1459} and
fermioinc
systems\cite{GW1441,KTT1429,GK150505856,FH160406527,KT170108264,WG170310937},
this result suggests that all 3+1D gapped liquid phases (such as SET and SPT
phases) for bosonic and fermionic systems with a finite unitary symmetry
(including trivial symmetry, \ie no symmetry) are classified by partially
gauging the symmetry of the bosonic/fermionic SPT orders\cite{LW180108530}.

However, the classification of  gapped non-liquid phases is still unclear (for
a review, see \Ref{PY200101722}).  In this paper, we are going to propose a
very systematic construction of 3+1D gapped non-liquid phases for bosonic and
fermionic systems with possible symmetry.  We hope our systematic construction
can lead to a classifying understanding of gapped non-liquid phases.  Our
construction is based on the above classification of gapped liquid phases and a
classification of gapped boundaries of those gapped liquid phases\cite{KW1458}.  

\begin{figure}[t]
\begin{center}
\includegraphics[scale=0.4]{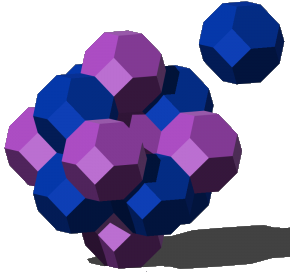} \end{center}
\caption{ 
The 3d space is divided into cells.  Each cell surface is occupied by a 2+1D
topological order. Each edge is occupied by an anomalous 1+1D topological
order, which is a gapped boundary of a stacking of the 2+1D topological orders.
}
\label{cell3d}
\end{figure}

In a simplified form of our construction, we divide the 3d space into cells
(see Fig. \ref{cell3d}), where only cell surfaces can overlap.  We put a 2+1D
topological order on each patch of overlapping surfaces.  The edges can be
viewed as a boundary of several stacked 2+1D topological orders. We put a
gapped boundary (a 1+1D anomalous topological order\cite{W1313,KW1458,K1467})
on each edge.  We refer to the constructed gapped states as \textbf{cellular
topological states}, to stress the intrinsic cellular structure (such as
foliation structure\cite{ZW1490,SC171205892,SC180310426}) in those gapped
states.  In general, cellular topological states are gapped non-liquid states,
although some special cellular topological states can be liquid states (\ie the
cellular structure disappears).

Our construction is similar to some constructions of SPT and SET
phases\cite{CLV1301,SH160408151,HH170509243,SQ181011013}.  It is also similar
to the layer, cage-net, or string-membrane  constructions of fracton
phases\cite{MH170100747,SK170403870,V170100762,PH180604687,SW181201613,F190802257}.
However, there are some important differences, which allow us to construct new
fracton states with fractal excitations,\cite{H11011962} even starting from
2+1D $\Z_2$ topological order.

We like to mention that there are also gapless non-liquid phases. They include
Fermi liquid in 2+1D and above,\cite{SM1403} as well as some models with
emergent graviton-like excitations.\cite{X0643,GW0600,GW1290}

\section{A simple construction}

\subsection{The construction}

\begin{figure}[t]
\begin{center}
\includegraphics[scale=0.5]{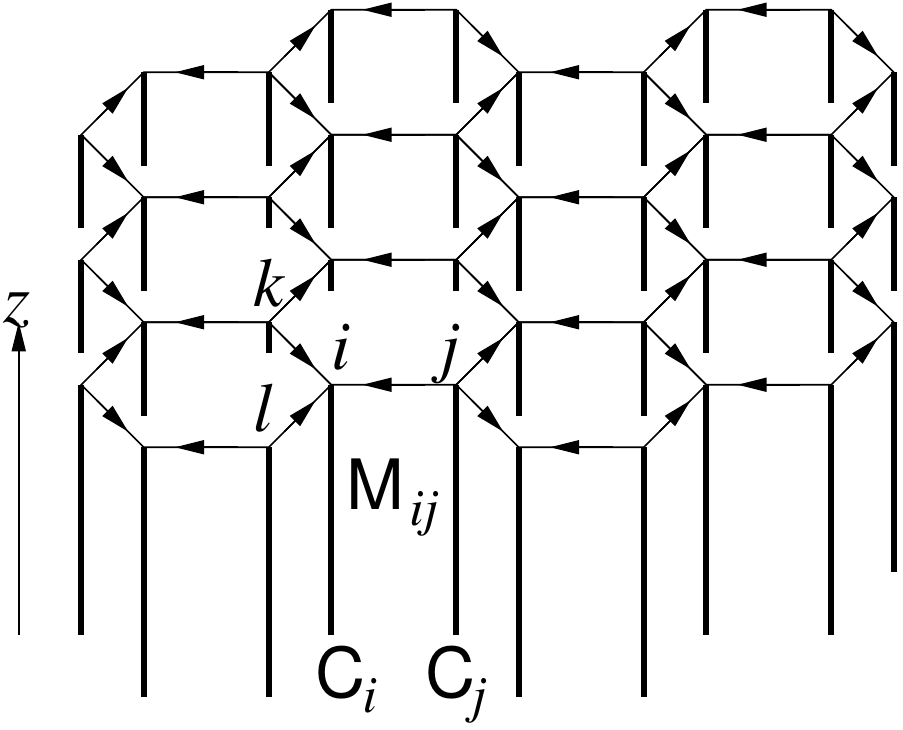} \end{center}
\caption{ 
The 3d space is decomposed into hexagonal column's.
The honeycomb lattice has two kinds of vertices:
type-A vertices have arrows pointing in, and
type-B vertices have arrows pointing out.
}
\label{hex3d}
\end{figure}

We first consider a very simple decomposition of the 3d space into hexagonal
column's
$ \R^3 = \cup_i (H_i\times \R_z)$,
where $H_i$ are non-overlaping hexagons whose union form the $x$-$y$ plane.
(see Fig. \ref{hex3d}).  We use $i,j$ to label the vertices and $ij$ the links
of honeycomb lattice in the  $x$-$y$ plane.  We then, put bosonic topological
orders $\sM_{ij}$ without symmetry on the faces of hexagonal column's
$\<ij\>\times \R_z$.

We note that the vertical line at the vertex $i$ of the honeycomb lattice is
the boundary of the topological order $\sM_{ij}\boxtimes \sM_{ik}\boxtimes
\sM_{il}$.  (Here $\sC\boxtimes \sD$ is the topological order obtained by
stacking topological orders $\sC$ and $\sD$.)  So in general, we can put a 1+1D
anomalous topological order\cite{W1313,KW1458,K1467} on the vertical line $i$ which
is described by a fusion category $\sC_i$\cite{LW1384}.  Those fusion
categories satisfy
\begin{align}
\label{CMh}
 \Bulk(\sC_i) &= 
\begin{cases}
\sM_i, & \text{ if $i$ is type-A},\\
\overline \sM_i, & \text{ if $i$ is type-B},\\
\end{cases}
\nonumber\\
\sM_i &= \sM_{ij} \boxtimes \sM_{ik} \boxtimes \sM_{il},
\end{align}
which means that 1+1D anomalous topological order $\sC_i$ is a gapped boundary
of 2+1D topological order $\sM_i$ or
$\overline\sM_i$\cite{KW1458,LWW1414,HW170600650,LW191108470}.  Here $\Bulk$ is
the holographic map that map a boundary anomalous topological order to its
unique corresponding bulk topological
order.\cite{KW1458,KZ150201690,KZ170200673,KZ200514178} $\Bulk$ is closely
related to the Drinfeld center, whose physical calculation is presented in
\Ref{LW1384}.  Also, the bar means the time reversal conjugate.

We see that, using the data $(\sM_{ij},\sC_i)$, we can construct a 3+1D gapped
phases for bosonic systems, which can be a non-liquid gapped phase.  In general
$\sM_{ij}$ can be non-Abelian topological orders.

\begin{figure}[t]
\begin{center}
\includegraphics[scale=0.45]{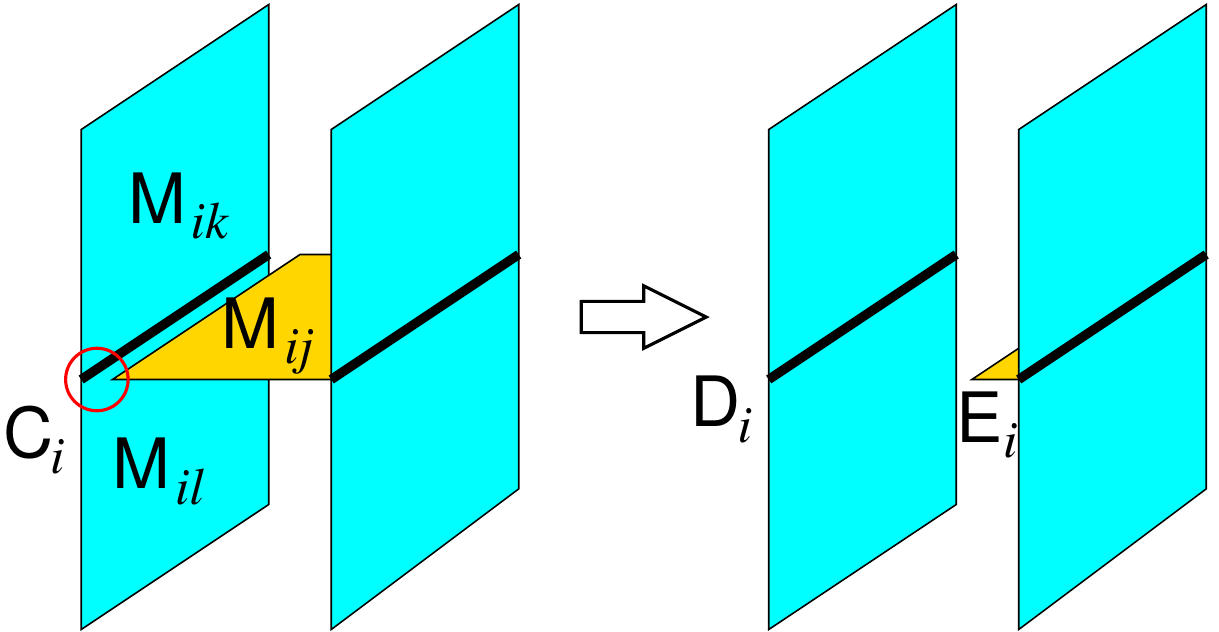} \end{center}
\caption{ 
For $\sC_i$ satisfying \eqn{CDE}, the layer $\sM_{ij}$ is detached.
}
\label{detach}
\end{figure}

We like to mention that if, for example, $\sC_i$ has a form
\begin{align}
\label{CDE}
\sC_i &= \sD_i \boxtimes \sE_i, 
\nonumber\\
\Bulk(\sE_i)  &= \sM_{ij}, \ \ \
\Bulk(\sD_i)  = \sM_{ik} \boxtimes \sM_{il},
\end{align}
then the layer $\sM_{ij}$ is not connected to the line at vertex-$i$.  Such a
layer can shrink to the line at vertex-$j$ on the other side (see Fig.
\ref{detach}), and hence can be removed (or correspond to trivial $\sM_{ij}$
case).  Thus we are looking for the so called entangled solutions of \eqn{CMh}
that do not have the form \eq{CDE}.
We also like to mention that the gapped boundaries of a 2+1D topological order
can be constructed via anyon condensation and are classified by the Lagrangian
algebra of the 2+1D topological order
\cite{KS10080654,KK11045047,WW1263,K13078244,HW13084673,HW14080014}. 

\subsection{Entanglement structure}

\begin{figure}[t]
\begin{center}
\includegraphics[scale=0.45]{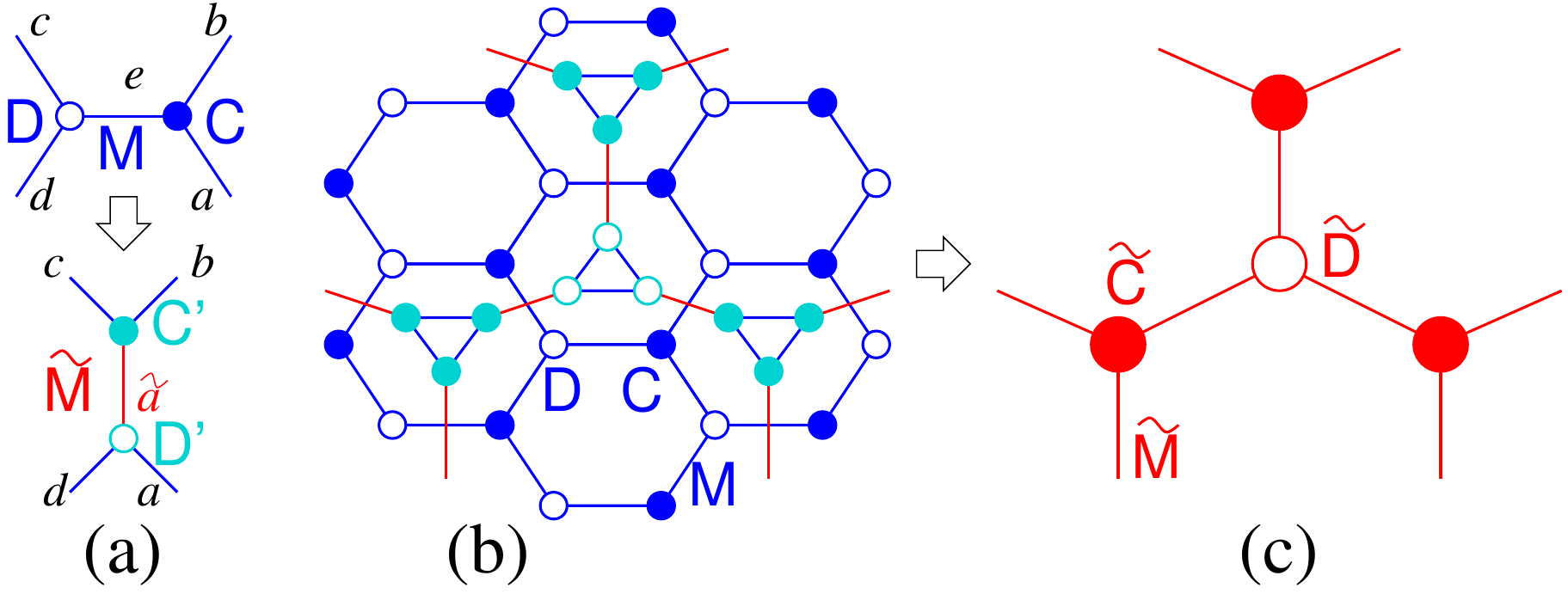} \end{center}
\caption{ 
(a) A deformation step $\sC \boxtimes_{\sM} \sD = \sC' \boxtimes_{\t\sM}
\sD'$ (see \eqn{deform}).  (b) Using the deformation step, we can change the
blue-hexagonal tensor network to the one formed by red links and light-blue
dots.  (c) Shrinking the triangles to the red dots produces the blue-hexagonal
tensor network (see \eqn{shrink}).  This completes a renormalization step
$(\sM,\sC,\sD) \to (\t\sM,\t\sC,\t\sD)$.
}
\label{RGhex}
\end{figure}

To understand the cellular topological state in Fig. \ref{hex3d} better, we
like to study the entanglement structure of the cellular topological state.
The entanglement structure can be revealed by the renormalization of the state
(see Fig. \ref{RGhex}).  The renormalization is done via a basic deformation
step in Fig.  \ref{RGhex}a, where fusing two boundaries $\sC,\sD$ and fusing
two boundaries $\sC',\sD'$ given rise to the same boundary of the four stacked
2+1D topological orders (described by the four outer lines): $\sC
\boxtimes_{\sM} \sD = \sC' \boxtimes_{\t\sM} \sD'$.

To describe such a deformation step more explicitly, we need a quantitative
description of the 2+1D topological order $\sM_{ij}$ and the 1+1D anomalous
topological order $\sC_i$.  The topological orders can be characterized by the
representations of mapping class groups for all Riemannian
surfaces.\cite{W9039,KW9327}  Here for
simplicity,\cite{MS170802796,BW180505736,WW190810381} we will only use the
representation for mapping class group of a torus.\cite{RSW0777,W150605768}  In
other words, we will use the $S_a^b,T_a^b$ matrices (the generators of a
modular representation of $SL(2,\Z)$) to characterize a 2+1D topological order
$\sM$, where $a,b$ label the types of the topological excitations in the
topological order.  Similarly, the gapped domain walls $\sC$ between two
topological orders characterized by $(S,T)$ and $(S',T')$ are characterized by
the wave function overlap of the degenerate ground states, $|\psi_{a}\>$ and
$|\psi'_{a'}\>$, of the two topological orders on torus:\cite{LW191108470}
\begin{align}
\label{olap}
\<\psi'_{a'}|\ee^{- H_W}|\psi_{a}\> = \ee^{-\si A_{T^2}+o(\frac{1}{A_{T^2}})}
C_{a'}^a 
\end{align} 
where $H_W$ is local hermitian operator like a Hamiltonian of a quantum system,
$A_{T^2}$ is the area of the torus $T^2$ and $C_{a'}^a$ is a topological
invariant that characterize the domain between the two topological orders.
$C_{a'}^a$ turns out to be non-negative integers for torus,
which satisfy\cite{LWW1414,LW191108470}
\begin{align}
\sum_{b'} {S'}_{a'}^{b'} C_{b'}^b &= \sum_a C_{a'}^a S_a^b, \ \ \ \ 
\sum_{b'} {T'}_{a'}^{b'} C_{b'}^b = \sum_a C_{a'}^a T_a^b, 
\nonumber\\
C_{a'}^a C_{b'}^b &\leq \sum_{c',c} N^{a'b'}_{c'} C_{c'}^c N^{ab}_c.
\end{align}
where $N^{ab}_c$ and $N^{a'b'}_{c'}$ are the fusion coefficients for the
topological excitations in the two topological orders. 

\begin{figure}[t]
\begin{center}
\includegraphics[scale=0.35]{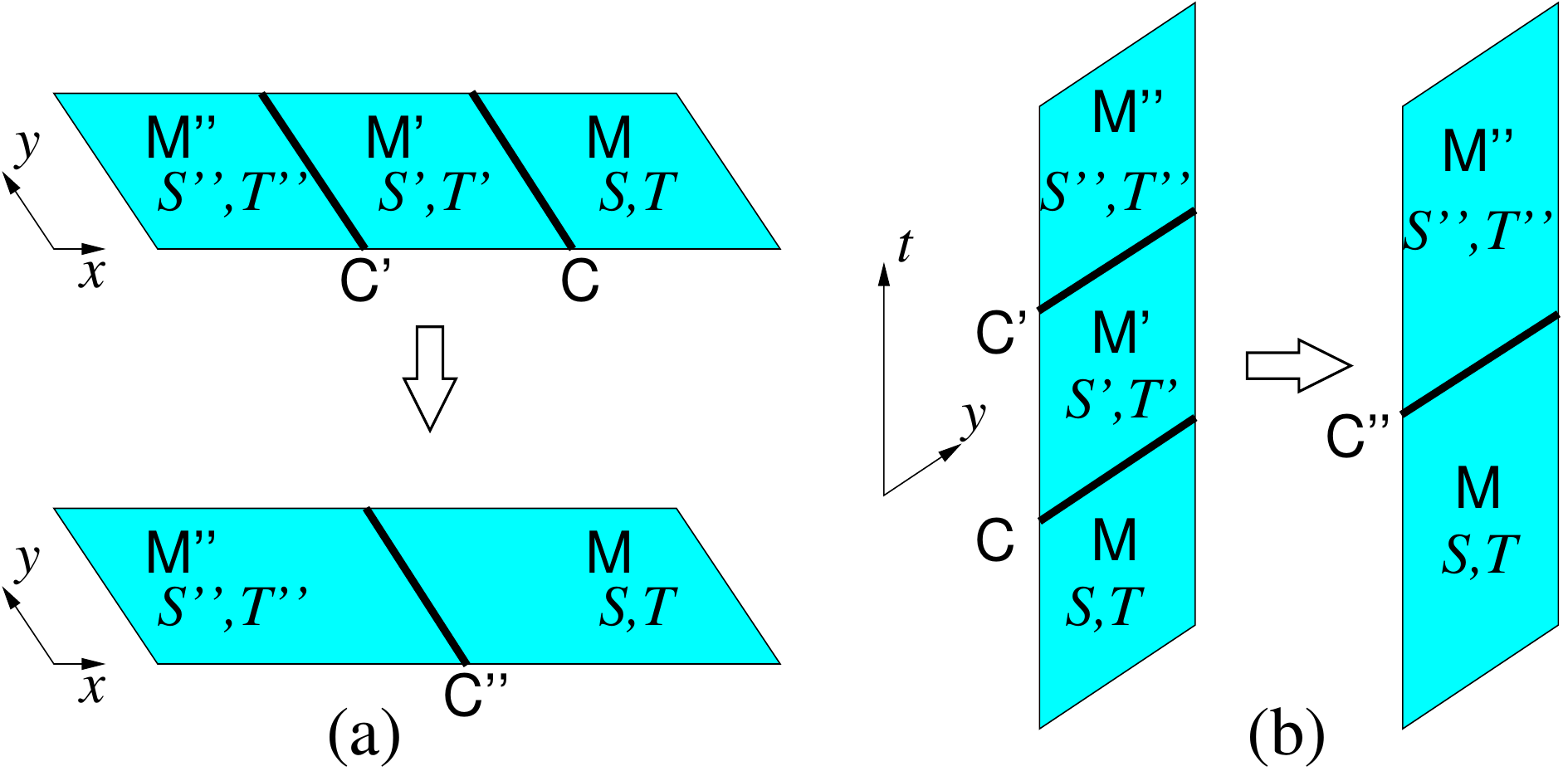} \end{center}
\caption{ 
(a) Fusion of two 1+1D domain walls $\sC$ and $\sC'$ 
connected by a 2+1D topological order $\sM'$ gives rise to
$\sC''=\sC'\boxtimes_{\sM'} \sC$.
The loop-like $t$-direction is not shown.
(b) Exchanging $x$ and $t$, we get the corresponding wave function overlapes.
Two wave function overlaps $C$ and $C'$ can be reduced to one
wave function overlap $C''$.
The loop-like $x$-direction is not shown.
}
\label{dwall}
\end{figure}

Now let us describe an elementary deformation step.
Consider three topological orders $\sM$, $\sM'$ and $\sM''$ characterized
by $(S,T)$, $(S',T')$, and $(S'',T'')$.  A tensor $C$ describes a domain wall
$\sC$ between $(S,T)$ and $(S',T')$, and a tensor $C'$ describes a domain wall
$\sC'$ between $(S',T')$ and $(S'',T'')$.  The two domain wall $\sC$ and $\sC'$
can fuse into a single domain wall $\sC''$ (see Fig. \ref{dwall}a):
\begin{align}
 \sC''= \sC' \boxtimes_{\sM'} \sC.
\end{align}
Note the $\sC$ and $\sC'$ are fused with a ``glue'' $\sM'$ (see Fig.
\ref{dwall})\cite{KZ150201690,KZ170200673}, which is indicated by the subscript
of $\boxtimes$.  It turns out that the  domain wall $\sC''$ is characterized by
a tensor $C''$
\begin{align}
 (C'')_{a''}^a = \sum_{a'} (C')_{a''}^{a'}C_{a'}^a, \ \text{ or } \
C''=C'C.
\end{align}
The above just describes the composition of wavefunction overlap
in Fig. \ref{dwall}b.

We note that the above elementary step is reversible, which can fuse two domain
walls or split a single domain wall.  A fusion followed a split in a different
direction produces the  elementary deformation step in Fig. \ref{RGhex}a.

If one side of the domain wall between $\sM$ and $\sM'$ is trivial (say $\sM'$
is trivial), then the domain wall (\ie the boundary of $\sM$) is described by $
C_\one^a \equiv C^a$ (or by $ C^\one_a \equiv C_a$ if the boundary is at the
opposite side of $\sM$, where $\one$ corresponds to the trivial excitation).
We see that the boundary $\sC_i$ in our construction (see Fig. \ref{hex3d}) is
characterized by non-negative integer tensor
\begin{align}
 \sC_i &\sim
\begin{cases}
C_{ a_{ij} a_{ik} a_{il} } , & \text{ if $i$ is type-A},\\
C^{ a_{ij} a_{ik} a_{il} } , & \text{ if $i$ is type-B},\\
\end{cases}
\end{align}
where $(a_{ij}, a_{ik}, a_{il})$ labels the topological excitations in $\sM_i =
\sM_{ij} \boxtimes \sM_{ik} \boxtimes \sM_{il}$ and $a_{ij}$ labels the
topological excitations in $\sM_{ij}$ \etc.

The above discussion suggests that we can view the blue honeycomb lattice in
Fig.  \ref{RGhex} as a tensor network, where the tensors at the solid-blue
vertices are given by $C^{ a_{ij} a_{ik} a_{il} }$, while the tensors at the
open-blue vertices are given by $D_{ a_{ij} a_{ik} a_{il} }$.  The link
$\<ij\>$ carries the index $a_{ij}$ which label the types of topological
excitations in $\sM_{ij}$.  \emph{The trace of the tensor network give us the
partition function, which is the ground state degeneracy of the cellular
topological state.}\cite{WW1263,LWW1414,LW191108470}

\begin{figure}[t]
\begin{center}
\includegraphics[scale=0.7]{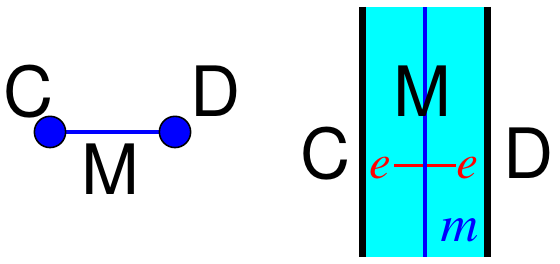} \end{center}
\caption{ 
A simple tensor network formed by two vertices connected by a link.  The link
corresponds to a 2+1D topological order $\sM$.  The vertices corresponds to a
1+1D anomalous topological order $\sC,\sD$.
}
\label{MCD}
\end{figure}

To see why trace of  tensor network give rise to ground state degeneracy, let
us consider a simple tensor network with two vertices connected by a
link\cite{WW1263,LWW1414}.  The link corresponds to a $\Z_2$ topological order
$\sM =\sGT^{2+1}_{\Z_2}$ (\ie the 2+1D $\Z_2$ gauge theory)\cite{RS9173,W9164}.
The $\Z_2$ topological order $\sGT^{2+1}_{\Z_2}$ has four type of topological
excitations $\one,e,m,f$, labeled by $a=1,2,3,4$ respectively.  The $S,T$
modular matrices are given by
\begin{align} 
\label{Z2ST}
  T&=
\begin{pmatrix}
    1&0&0&0\\
    0&1&0&0\\
    0&0&1&0\\
    0&0&0&-1
  \end{pmatrix}, 
&
  S&=\frac12 \begin{pmatrix}
    1&1&1&1\\
    1&1&-1&-1\\
    1&-1&1&-1\\
    1&-1&-1&1
  \end{pmatrix}.
\end{align}
The  $\Z_2$ topological order has two gapped boundayies: $\sC_e$ from $e$-particle
condensation and $\sC_m$ from $m$-particle condensation\cite{KK11045047}.  There
are described by the following rank-1 tensors ($a=1,\cdots,4$)
\begin{align}
 \sC_e:\ (C_e^a) = \begin{pmatrix}
    1\\
    1\\
    0\\
    0\\
  \end{pmatrix}
,\ \ \ \
\sC_m:\ (C_m^a) = \begin{pmatrix}
    1\\
    0\\
    1\\
    0\\
  \end{pmatrix}
\end{align}
If both  boundaries in Fig. \ref{MCD} are given by $\sC=\sD=\sC_e$, then the
ground state degeneracy of the system is given by $\sum_a C_e^a C_e^a =2$.  This
result can also be obtained using the $e$-string operator $W_e$ that creates a
pairs of $e$-particle at its ends, and the $m$-string operator $W_m$ that
creates a pairs of $m$-particle at its ends.  Since $e$-particles condense at
the boundaries, the open $e$-string operator, $W_e$, connecting the two
boundaries does change the energy (\ie commute with the Hamiltonian).  A loop
of the $m$-string operator in $z$-direction, $W_m^z$, also commute with the
Hamiltonian. Since the $e$-string operator and the $m$-string operator
intersect at one point and anti-commute $W_e W_m^z = - W_m^z W_e$, the ground
states are 2-fold degenerate.

If the boundaries in Fig. \ref{MCD} are given by $\sC=\sC_e$ and $\sC=\sC_m$,
then the ground state degeneracy of the system is given by $\sum_a C_e^a C_m^a
=1$. In this case, there is no string operators that connect the two boundaries
and create two condensing particles.

With the above tensor representation of the boundaries, the renormalization of
the cellular topological state becomes the standard renormalization of tensor
network.\cite{VC0466,LN0701} Let us assume that, in the hexagonal tensor
network (see Fig. \ref{hex3d} and \ref{RGhex}b), all $\sM_{ij}$ are the same
$\sM_{ij}=\sM$, whose topological excitations are labeled by $a,b,c,\cdots$.
The boundaries $\sC_i$ at the solid-bule vertices are given by $\sC_i=\sC$ (or
by tensor $C^{abc}$), while boundaries $\sC_i$ at the open-bule vertices are
given by $\sC_i=\sD$ (or by tensor $D_{abc}$).  For simplicity, we will assume
\begin{align}
\label{csymm}
 C^{abc} = C^{cab}, \ \ \ \ D_{abc} = D_{cab}.
\end{align}
Then, the deformation in Fig. \ref{RGhex}a
is explicitly given by the following tensor relation:
\begin{align}
\label{deform}
\sum_e  C^{eab} D_{ecd} = \sum_{\t a} (C')^{\t abc} (D')_{\t ada} .
\end{align}
where $\t a$ label the topological excitations in a new 2+1D topological order
$\t\sM$.  We like to mention that the deformation \eq{deform} is not unique.
There can be many choices of $\t\sM,\sC',\sD'$ that satisfy \eqn{deform}.  We
like to find the deformation where $\t\sM$ has minimal total quantum dimension
$D=\sqrt{\sum_{\t a} d^2_{\t a}}$. Here $d_{\t a}$ is the quantum dimensions of
topological excitations in $\t\sM$.  Later, we will see that if the resulting
$\t \sM$ is trivial or equal to the original $\sM$, then the corresponding
cellular topological state may be a liquid state.

We like to remark that, as we will see later, a cellular state contains extra
local structures that are not related to the universal class of a gapped state.
So, by choosing $\t \sM$ to have minimal total quantum dimension, we hope to
obtain the simplest cellular topological state after each step of
renormalization, trying to remove those local structures as much as possible.

We can use the deformation Fig. \ref{RGhex}a to deform the blue hexagonal
tensor network in  Fig. \ref{RGhex}b to the one described by red links in  Fig.
\ref{RGhex}b.  We then shrink the small triangles in  Fig. \ref{RGhex}b to a
point and obtain a new red  hexagonal tensor network in Fig. \ref{RGhex}c.
The new boundaries $\t \sC$ and $\t \sD$ are given by
\begin{align}
\label{shrink}
 \t C^{\t a\t b\t c} = \sum_{a,b,c} (C')^{\t acb} (C')^{\t bac} (C')^{\t cba} ,
\nonumber\\
 \t D_{\t a\t b\t c} = \sum_{a,b,c} (D')_{\t acb} (D')_{\t bac} (D')_{\t cba} .
\end{align}
The two relations \eq{deform} and \eq{shrink} define the renormalization
of the cellular topological state.

\section{Cellular topological states from  2+1D $\Z_2$ topological order}

\subsection{A general construction}

In this section, we are going to construct some simple cellular topological
states in Fig. \ref{hex3d} by choosing $\sM_{ij}$ to be the same 2+1D $\Z_2$
topological order $\sM_{ij} = \sGT^{2+1}_{\Z_2}$. We find that 
$ \sGT^{2+1}_{\Z_2}\boxtimes \sGT^{2+1}_{\Z_2}\boxtimes \sGT^{2+1}_{\Z_2} $ has 10 types of gapped
boundaries, $\Bulk(\sC_i) = \sGT^{2+1}_{\Z_2}\boxtimes \sGT^{2+1}_{\Z_2}\boxtimes \sGT^{2+1}_{\Z_2} $, that
are entangled (\ie do not have the form in \eqn{CDE}).  Their tensor
representations, $C_i^{abc}$, are givein by (only non-zero elements are
listed):
\begin{align}
\label{Cs}
\sC_{1} &: C_{1}^{111},C_{1}^{122},C_{1}^{212},C_{1}^{221},C_{1}^{333},C_{1}^{344},C_{1}^{434},C_{1}^{443}=1, 
\nonumber\\
\sC_{2} &: C_{2}^{111},C_{2}^{144},C_{2}^{223},C_{2}^{232},C_{2}^{322},C_{2}^{333},C_{2}^{414},C_{2}^{441}=1, 
\nonumber\\
\sC_{3} &: C_{3}^{111},C_{3}^{133},C_{3}^{222},C_{3}^{244},C_{3}^{313},C_{3}^{331},C_{3}^{424},C_{3}^{442}=1, 
\nonumber\\
\sC_{4} &: C_{4}^{111},C_{4}^{144},C_{4}^{222},C_{4}^{233},C_{4}^{323},C_{4}^{332},C_{4}^{414},C_{4}^{441}=1, 
\nonumber\\
\sC_{5} &: C_{5}^{111},C_{5}^{132},C_{5}^{212},C_{5}^{231},C_{5}^{323},C_{5}^{344},C_{5}^{424},C_{5}^{443}=1, 
\nonumber\\
\sC_{6} &: C_{6}^{111},C_{6}^{132},C_{6}^{223},C_{6}^{244},C_{6}^{312},C_{6}^{331},C_{6}^{424},C_{6}^{443}=1, 
\nonumber\\
\sC_{7} &: C_{7}^{111},C_{7}^{123},C_{7}^{213},C_{7}^{221},C_{7}^{332},C_{7}^{344},C_{7}^{434},C_{7}^{442}=1, 
\nonumber\\
\sC_{8} &: C_{8}^{111},C_{8}^{123},C_{8}^{232},C_{8}^{244},C_{8}^{313},C_{8}^{321},C_{8}^{434},C_{8}^{442}=1, 
\nonumber\\
\sC_{9} &: C_{9}^{111},C_{9}^{133},C_{9}^{213},C_{9}^{231},C_{9}^{322},C_{9}^{344},C_{9}^{424},C_{9}^{442}=1, 
\nonumber\\
\sC_{10} &: C_{10}^{111},C_{10}^{122},C_{10}^{233},C_{10}^{244},C_{10}^{312},C_{10}^{321},C_{10}^{434},C_{10}^{443}=1. 
\end{align}

The first four, $( \sC_1, \sC_2, \sC_3, \sC_4)$, are cyclic symmetric
\eq{csymm}.  Let us examine those four types of the boundaries of
$\sGT^{2+1}_{\Z_2}\boxtimes \sGT^{2+1}_{\Z_2}\boxtimes \sGT^{2+1}_{\Z_2} $ in more details. First we note
that $\sC_1$ and $\sC_3$ as well as $\sC_2$ and $\sC_4$ differ by an
automorphism of the $\Z_2$ topological order: $e_i \leftrightarrow m_i$, where
$i=1,2,3$ labels the three $\Z_2$ topological orders in $\sGT^{2+1}_{\Z_2}\boxtimes
\sGT^{2+1}_{\Z_2}\boxtimes \sGT^{2+1}_{\Z_2} $.  Those boundaries are formed by condensing the
topological excitations in the three $\Z_2$ topological orders\cite{KK11045047,K13078244,HW13084673,HW14080014}.
In the following, we list the condensing excitations (the generators)
for the four boundaries:
\begin{align}
\sC_{1} &:\ \ {e_2 e_3}, && {e_1  e_3}, && {e_1 e_2 }, && {m_1 m_2 m_3};
\nonumber\\
\sC_{2} &:\ \ {f_1 f_2 }, && {f_1 f_3}, && {f_2 f_3}, && {m_1 m_2 m_3};
\nonumber\\
\sC_{3} &:\ \ {m_2 m_3}, && {m_1 m_3}, && {m_1 m_2}, && {e_1 e_2  e_3};
\nonumber\\
\sC_{4} &:\ \ {f_2 f_3}, && {f_1 f_3}, && {f_1 f_2}, && {e_1 e_2  e_3};
\end{align}
which are obtained from the tensor indices $abc$ with $C^{abc}=1$.

\begin{figure}[t]
\begin{center}
\includegraphics[scale=0.5]{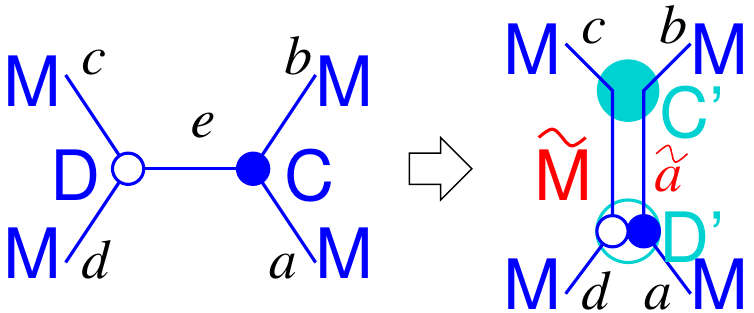} \end{center}
\caption{
If we choose $\t\sM = \sM\boxtimes \sM$, then
\eqn{deform} always has solutions.
}
\label{sol}
\end{figure}

If we assume $\t\sM$ to be the trival topological order, then the equation
\eqn{deform} for the deformation in Fig. \ref{deform}a has no solutions, for
those cyclic symmetric boundaries.  If we assume $\t\sM$ to be given by
the 2+1D $\Z_2$ topological order $\sGT^{2+1}_{\Z_2}$, then for the following
$(\sC,\sD)$'s
\begin{align}
(\sC_{1}, \sC_{2}),\ (\sC_{1}, \sC_{3}),\ (\sC_{1}, \sC_{4}),\ (\sC_{2}, \sC_{3}),\ (\sC_{3}, \sC_{4}), 
\end{align}
the deformation \eqn{deform} also has no solution.  But, for other cyclic
symmetric boundaries $\sC,\sD$'s, the deformation \eqn{deform} has two
solutions, which are given by (see Fig. \ref{deform})
\begin{align}
(\sC,\sD)=(\sC_{1}, \sC_{1}) &\to (\sC',\sD')= (\sC_{1},\sC_{1}) \text{ or }  (\sC_{10},\sC_{10}), 
\nonumber\\
(\sC,\sD)=(\sC_{3}, \sC_{3}) &\to (\sC',\sD')= (\sC_{3},\sC_{3}) \text{ or }  (\sC_{9},\sC_{9}), 
\nonumber\\
(\sC,\sD)=(\sC_{2}, \sC_{2}) &\to (\sC',\sD')= (\sC_{2},\sC_{2}) \text{ or }  (\sC_{4},\sC_{4}), 
\nonumber\\
(\sC,\sD)=(\sC_{4}, \sC_{4}) &\to (\sC',\sD')= (\sC_{2},\sC_{2}) \text{ or }  (\sC_{4},\sC_{4}), 
\nonumber\\
(\sC,\sD)=(\sC_{2}, \sC_{4}) &\to (\sC',\sD')= (\sC_{2},\sC_{4}) \text{ or }  (\sC_{4},\sC_{2}). 
\end{align}
If we choose $\t\sM$ to be a more general topological order,
such as $\t\sM = \sM\boxtimes \sM$, then
\eqn{deform} always has solutions (see Fig. \ref{sol}).

After obtaining $\sC'$ and $\sD'$, we can perform the shrinking operation
\eq{shrink} (see Fig. \ref{RGhex}) to obtain $\t \sC,\t \sD$:
\begin{align}
\label{CtC}
(\sC,\sD)=(\sC_{1}, \sC_{1}) &\to (\t\sC,\t\sD)= (2\sC_{1}, 2\sC_{1})\text{ or } (2\sC_{3}, 2\sC_{3}), 
\nonumber\\
(\sC,\sD)=(\sC_{3}, \sC_{3}) &\to (\t\sC,\t\sD)= (2\sC_{3}, 2\sC_{3})\text{ or } (2\sC_{1}, 2\sC_{1}), 
\nonumber\\
(\sC,\sD)=(\sC_{2}, \sC_{2}) &\to (\t\sC,\t\sD)= (2\sC_{2}, 2\sC_{2})\text{ or } (2\sC_{4}, 2\sC_{4}), 
\nonumber\\
(\sC,\sD)=(\sC_{4}, \sC_{4}) &\to (\t\sC,\t\sD)= (2\sC_{2}, 2\sC_{2})\text{ or } (2\sC_{4}, 2\sC_{4}), 
\nonumber\\
(\sC,\sD)=(\sC_{2}, \sC_{4}) &\to (\t\sC,\t\sD)= (2\sC_{2}, 2\sC_{4})\text{ or } (2\sC_{4}, 2\sC_{2}). 
\end{align}
Here $2\sC \equiv \sC\oplus \sC$ means that the boundary is formed by
accidentally degenerate $\sC$ and $\sC$.  Since $\t\sC$ comes from fusing three
$\sC$'s. We roughly have a fusion rule for the boundaries: $ \sC\boxtimes
\sC\boxtimes \sC\sim \t\sC$.  The results \eq{CtC} suggest that the boundary
$\sC$ and $\sD$ have a quantum dimension $\sqrt 2$.  So the ground state
degeneracy is roughly given by $2^{\frac{N_A+N_B}{2}}$ (up to a finite factor),
where $N_A$ and $N_B$ are the number of type-A and type-B vertices (see Fig.
\ref{hex3d}).  In other words the ground state degeneracy is roughly given by
$2^{N_h}$  where $N_h$ is the number of the hexagons (see Fig. \ref{hex3d}).

In our above discussions, we have assumed that the vertices in the honeycomb
lattice (see Fig. \ref{hex3d}) is far apart.  This leads to the accidentally
degeneracy of two $\sC$'s.  However, in reality,  the vertices in the honeycomb
lattice have a small separation.  In this case, the degeneracy  of two $\sC$'s
is split.  

To summarize, we constructed five cellular topological phases labeled by the
following $(\sM,\sC,\sD)$'s:
\begin{align}
&(\sGT^{2+1}_{\Z_2},\sC_1,\sC_1),\
 (\sGT^{2+1}_{\Z_2},\sC_2,\sC_2),\
 (\sGT^{2+1}_{\Z_2},\sC_3,\sC_3),
\nonumber\\
&(\sGT^{2+1}_{\Z_2},\sC_4,\sC_4),\
 (\sGT^{2+1}_{\Z_2},\sC_2,\sC_4).
\end{align}
Those phases have the key properties that under the renormalization
$(\sM,\sC,\sD) \to(\t\sM,\t\sC,\t\sD)$ in Fig.  \ref{RGhex}, we cannot reduce
the 2+1D topological order $\sM$ to the trivial one, but $\sM$ can be unchanged
under renormalization: $\sM=\sGT^{2+1}_{\Z_2} \to \t\sM=\sGT^{2+1}_{\Z_2}$.  Later,
we will see that the invariance of $\sM$ under renormalization suggests that
the corresponding cellular topological state is a liquid state.

We also constructed five cellular topological phases labeled by the
following $(\sM,\sC,\sD)$'s:
\begin{align}
&(\sGT^{2+1}_{\Z_2},\sC_1,\sC_2),\
 (\sGT^{2+1}_{\Z_2},\sC_1,\sC_3),\
 (\sGT^{2+1}_{\Z_2},\sC_1,\sC_4),
\nonumber\\
&(\sGT^{2+1}_{\Z_2},\sC_2,\sC_3),\
 (\sGT^{2+1}_{\Z_2},\sC_3,\sC_4).
\end{align}
Those phases have the key properties that under the renormalization
$(\sM,\sC,\sD) \to(\t\sM,\t\sC,\t\sD)$ in Fig.  \ref{RGhex}, we cannot reduce
the 2+1D topological order $\sM$ to the trivial one, and $\sM$ cannot be
unchanged under the renormalization.  Later, we will see that the
non-invariance of $\sM$ under renormalization suggests that the corresponding
cellular topological state is a non-liquid state.

\subsection{Cellular topological state $(\sGT^{2+1}_{\Z_2},\sC_1,\sC_1)$}

Some cellular topological states are non-liquid states, while other cellular
topological states are actually liquid states.  In this section, we are going
to discuss a cellular topological state $(\sGT^{2+1}_{\Z_2},\sC_1,\sC_1)$, and
show that it is actually a gapped liquid state -- a 3+1D $\Z_2$ topological
ordered state $\sGT^{3+1}_{\Z_2}$ described by $\Z_2$ gauge theory.

The cellular topological state $(\sGT^{2+1}_{\Z_2},\sC_1,\sC_1)$ is constructed
using 2+1D $\Z_2$ topological order, and choosing the junction of three  $\Z_2$
topological orders to be the 1+1D anomalous topological order $\sC_1$ in
\eqn{Cs} (see Fig.  \ref{hex3d}).

\begin{figure}[t]
\begin{center}
\includegraphics[scale=0.5]{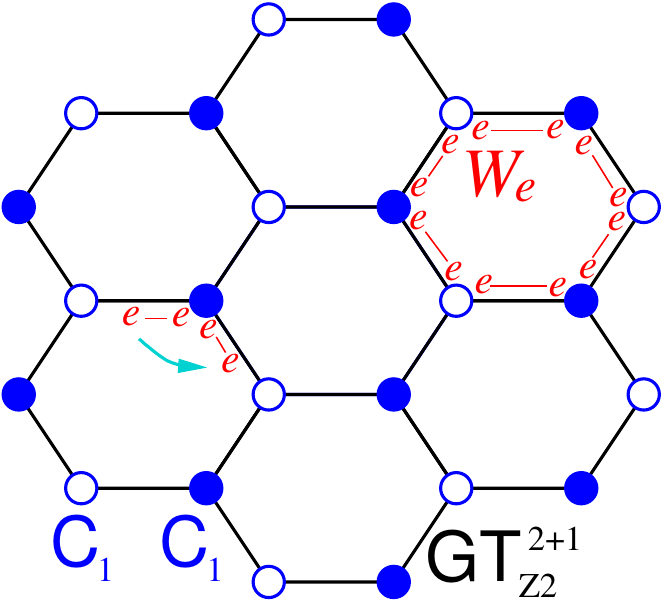} \end{center}
\caption{In the cellular topological state $(\sGT^{2+1}_{\Z_2},\sC_1,\sC_1)$, the
$e$-particle can move freely in 3d space.  The unmarked links are in sector-1.
A configuration with a loop of links in sector-2 (marked by $e-e$) corresponds
to another degenerate ground state.
}
\label{Z2C1C1}
\end{figure}

\begin{figure}[t]
\begin{center}
\includegraphics[scale=0.5]{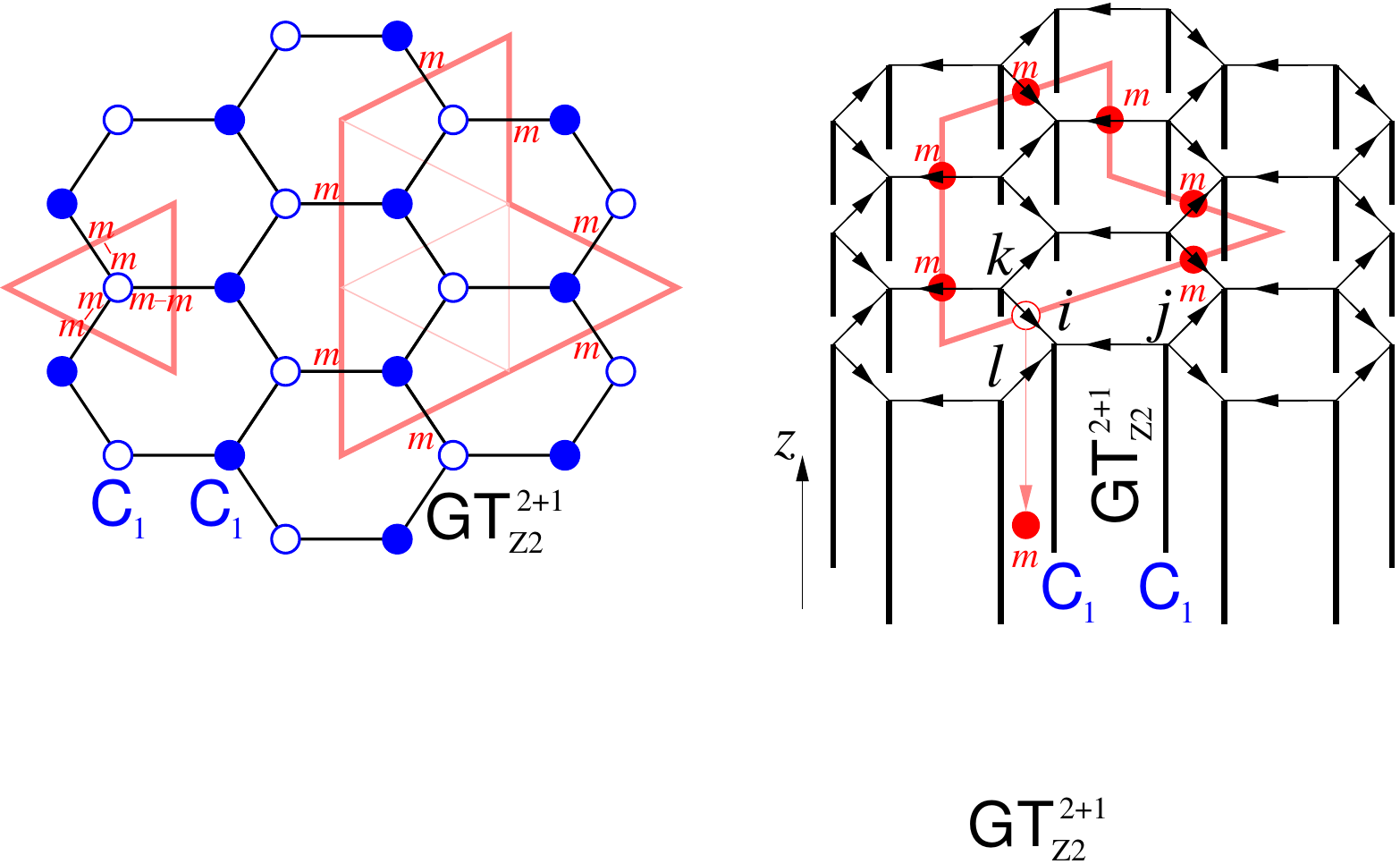} \end{center}
\caption{ 
In the cellular topological state $(\sGT^{2+1}_{\Z_2},\sC_1,\sC_1)$,
the $m$-particles must form a closed loop in the dual honeycomb lattice.
}
\label{Z2C1C1m}
\end{figure}

The 1+1D topological order $\sC_1$ has a condensation of $e_1e_2$, $e_2e_3$,
and $e_3e_1$, for the excisions in the connected 2+1D topological order. This
means that the $e$-particles can freely move between the 2+1D topological
orders $\sGT^{2+1}_{\Z_2}$ connected by the 1+1D topological order $\sC_1$ (see
Fig. \ref{Z2C1C1}).
In other words, the $e$-particle can move freely in the whole 3d space.

From the renormalization of the corresponding tensor network 
\begin{align}
 (\sC,\sD)=(\sC_{1}, \sC_{1}) &\to (\t\sC,\t\sD)= (2\sC_{1}, 2\sC_{1}),
\end{align}
we find that the ground state degeneracy is roughly given by $2^{N_h}$.  Such
degeneracy can be understood by using the closed $e$-string operator $W_e$ that
move an $e$-particle around a hexagon and the closed $m$-string operators
$W_m^z$ that wraps around in the $z$-direction (see Fig. \ref{MCD}).  Both
closed string operators commute with the Hamiltonian. Since $W_e$ and $W_m^z$
anti-commute when they intersects, we find that each hexagon contributes a
factor 2 (corresponding to $W_e=\pm1$) to the ground state degeneracy.

The  cellular topological state has a tensor network representation (see Fig.
\ref{hex3d} and \ref{RGhex}).  We can also compute ground state state
degeneracy using the trace of the tensor network.  Each link of the tensor
network has a label $a=1,2,3,4$.  The label $1$ corresponds to a stripe of
$\Z_2$ topological order in the trivial sector.  The label $2,3,4$ correspond to
a stripe in the non-trivial sectors.  Applying an open $e$-string operator
$W_e$ connecting the two boundaries to the trivial sector produces the
sector-$2$ (see Fig. \ref{MCD}).  Similarly, applying the open $m$-string
($f$-string) operator $W_m$ ($W_f$) connecting the two boundaries to the
trivial sector produces the sector-$3$ (the sector-$4$).

A ground state of the cellular topological phase is given by the stripes of
$\Z_2$ topological orders, all in the trivial sector (\ie with label $a=1$ on
all links).  Now we apply a loop $e$-string operators $W_e^\text{loop}$ 
on some links to make
them to be a small loop of sector-$2$ (see Fig. \ref{Z2C1C1}).  The
configuration corresponds to another degenerate  ground state.  Thus each
hexagon contributes a factor 2 to the ground state degeneracy.

When the separation between vertices is small, the operators  $W_e^\text{loop}$
are local operators.  We may include such operators in the Hamlitonian $\del H
= J \sum W_e^\text{loop}$.  The new Hamiltonian no longer commute with
$m$-string operators $W_m$.  So $\del H$ splits the ground state degeneracy.
The new ground states is believed to have a finite degeneracy independent of
system size.

The condensation $m_1m_2m_3$ at the 1+1D topological order $\sC_1$ implies that
we can create three $m$-particles on the neighboring three 2+1D $\Z_2$
topological orders, which form a small triangle in the dual honeycomb lattice.
Putting many small triangles together gives us a loop in the dual honeycomb
lattice formed by the $m$-particles (see Fig. \ref{Z2C1C1m}).  However, the
$z$-coordinates of the $m$-particles can be arbitrary.

But if we add the  $\del H = J \sum W_e^\text{loop}$ term to the Hamiltonian,
it will confine two $m$-particles in the same stripe of the $\Z_2$ topological
order.  In this case the above loop of the $m$-particles must have similar
$z$-coordinates, in order to reduce the energy.

Those properties suggest that the cellular topological state
$(\sGT^{2+1}_{\Z_2},\sC_1,\sC_1)$ is a 3+1D $\Z_2$ topological order
$\sGT_{\Z_2}^{3+1}$.  The free-moving $e$-particle is the point-like $\Z_2$-charge
in $\sGT_{\Z_2}^{3+1}$.  The loop of $m$-particles is the $\Z_2$-flux loop in
$\sGT_{\Z_2}^{3+1}$.

\subsection{Cellular topological state $(\sGT^{2+1}_{\Z_2},\sC_2,\sC_2)$}

\begin{figure}[t]
\begin{center}
\includegraphics[scale=0.5]{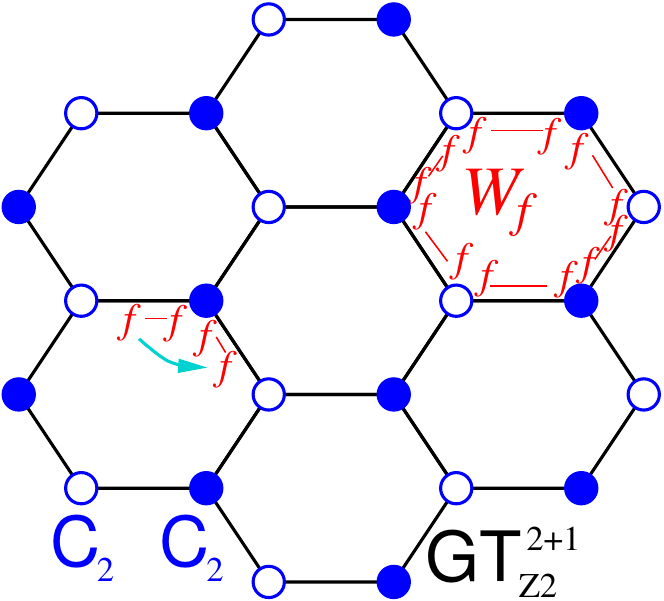} \end{center}
\caption{In the cellular topological state $(\sGT^{2+1}_{\Z_2},\sC_2,\sC_2)$,
the $f$-particle can move freely in 3d space.
}
\label{Z2C2C2}
\end{figure}

The cellular topological state $(\sGT^{2+1}_{\Z_2},\sC_2,\sC_2)$ is also a gapped
liquid state -- a 3+1D $\Z_2^f$ topological ordered state $\sGT^{3+1}_{\Z_2^f}$
described by twisted $\Z_2$ gauge theory where the point-like $\Z_2$-charge is a
fermion.\cite{LW0316}

The 1+1D topological order $\sC_2$ has a condensation of $f_1f_2$, $f_2f_3$,
and $f_3f_1$, for the excitations in the connected 2+1D topological orders.
This means that the $f$-particles can freely move between the 2+1D topological
orders $\sGT^{2+1}_{\Z_2}$ connected by the 1+1D topological order $\sC_2$ (see
Fig. \ref{Z2C2C2}).  In other words, the $f$-particle can move freely in the
whole 3d space, which corresponds to the point-like $\Z_2$-charge in the 3+1D
$\Z_2^f$ topological order $\sGT^{3+1}_{\Z_2^f}$.  Similarly, the loop of
$m$-particles is the $\Z_2$-flux loop in $\sGT_{\Z_2^f}^{3+1}$.

\begin{figure}[t]
\begin{center}
\includegraphics[scale=0.5]{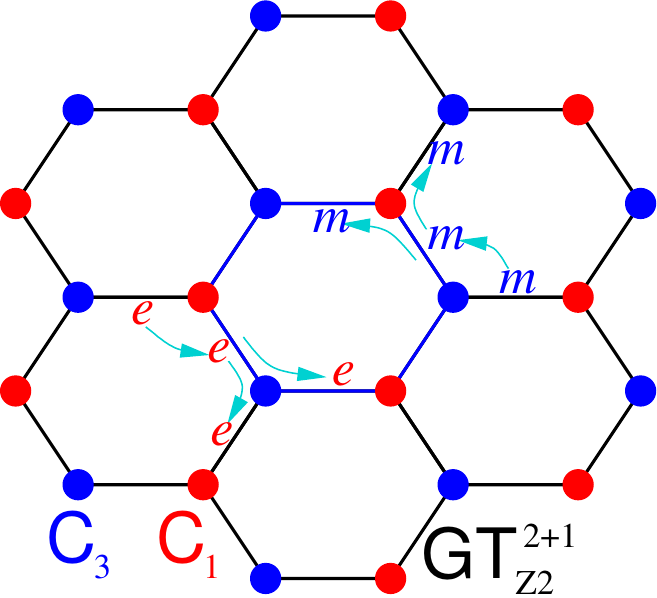} \end{center}
\caption{In the cellular topological state $(\sGT^{2+1}_{\Z_2},\sC_1,\sC_3)$,
the $e$-particle can move across the $\sC_1$ boundary,
and the $m$-particle can move across the $\sC_3$ boundary.
However, the long distant motion of $e$- and $m$-particles are blocked.
}
\label{Z2C1C3}
\end{figure}

\begin{figure}[t]
\begin{center}
\includegraphics[scale=0.5]{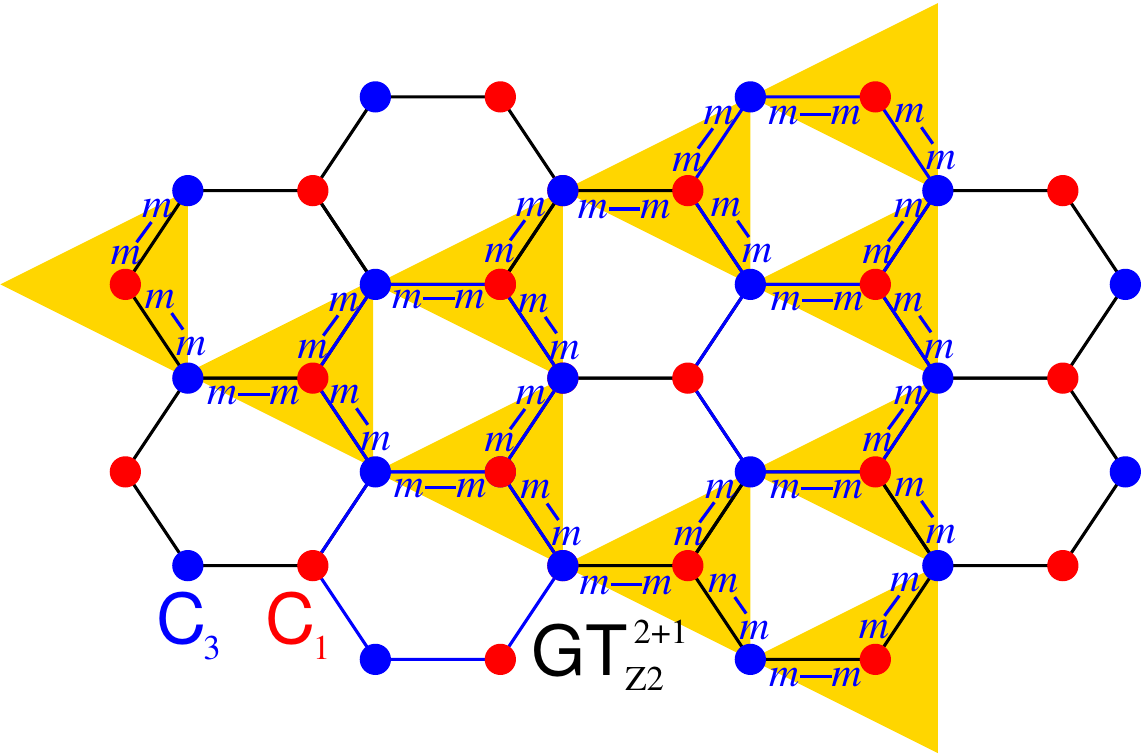} \end{center}
\caption{
A configuration in the ground state.  The unmarked links are in sector-1.  The
links marked by $m-m$ are in sector-3.
}
\label{Z2C1C3cnf}
\end{figure}

\subsection{Cellular topological state $(\sGT^{2+1}_{\Z_2},\sC_1,\sC_3)$}

The cellular topological state $(\sGT^{2+1}_{\Z_2},\sC_1,\sC_3)$ is a gapped
non-liquid state, which is a fracton state with fractal excitations.  In such a
cellular topological state, $e$-particle ($m$-particle) can move across the
$\sC_1$ ($\sC_3$) boundaries (see Fig.  \ref{Z2C1C3}).  But the motion of
$e$-particle ($m$-particle) is blocked by the $\sC_3$ ($\sC_1$) boundaries.  To
move across the  $\sC_3$ ($\sC_1$) boundaries, the $e$-particle ($m$-particle)
must split into two (see Fig. \ref{Z2C1C3}).  So the $e$- and $m$-particles
cannot move freely in $x$-$y$ direction, indicating that the cellular
topological state maybe a non-liquid state.  However, the $e$-particle and
$m$-particle can move freely in $z$-direction within a stripe of $\Z_2$
topological order (see Fig.  \ref{Z2C1C1m}).

Remember that a ground state of the cellular topological phase is given by the
stripes of $\Z_2$ topological orders, all in the trivial sector (\ie with label
$a=1$ on all links).  Now we apply the  $m$-string operators $W_m$ on some
links to make them to be the sector-$3$.  The created $m$-particle bound state
on a vertex must be able to condense on the boundary.  In this case, we create
another degenerate ground state (see Fig. \ref{Z2C1C3cnf}).  We can also  apply
the  $e$-string operators $W_e$ on some links to make them to be the
sector-$2$.  The created $e$-particle bound state on the boundary must be able
to condense on the boundary (the resulting configuration is similar to Fig.
\ref{Z2C1C3cnf}).  This way, we obtain another degenerate ground state.  We can
also apply the $e$-string and $m$-string operators together to obtain new
degenerate ground states.  Counting all such configurations give us the ground
state degeneracy.  We note that different degenerate ground states have a large
separation of code distance, which increases with system size.

From Fig. \ref{Z2C1C3cnf}, we see that, in the ground state, the links in
sector-$3$ form many small triangles. A corner of a triangle must connect to
one and only one  corner  of another triangle.  This way, the links in
sector-$3$ form a fractal (see Fig.  \ref{Z2C1C3cnf}).  This implies that the
cellular topological state $(\sGT^{2+1}_{\Z_2},\sC_1,\sC_3)$ is a 
fracton state with fractal excitations.

If a corner of a triangle is not connected to any corner triangle, such a
corner will represent a point-like excitation.  But the $x$-$y$ motion  for
such a point-like excitation is highly restricted,  like the point excitations
in Haah's cubic code.  Such kind of point excitations are called fractons.  We
see fractons are created at the corners of the fractal operator.  However,
fractons  can move freely in the $z$-direction within a stripe of $\Z_2$
topological order.

We believe that the cellular topological states, $
(\sGT^{2+1}_{\Z_2},\sC_1,\sC_2)$, $(\sGT^{2+1}_{\Z_2},\sC_1,\sC_4)$, $
(\sGT^{2+1}_{\Z_2},\sC_2,\sC_3)$, $(\sGT^{2+1}_{\Z_2},\sC_3,\sC_4)$, are similar
to the cellular topological states $ (\sGT^{2+1}_{\Z_2},\sC_1,\sC_3)$ discussed
above.  They should also be fracton states with fractal excitations.

\begin{figure}[t]
\begin{center}
\includegraphics[scale=0.5]{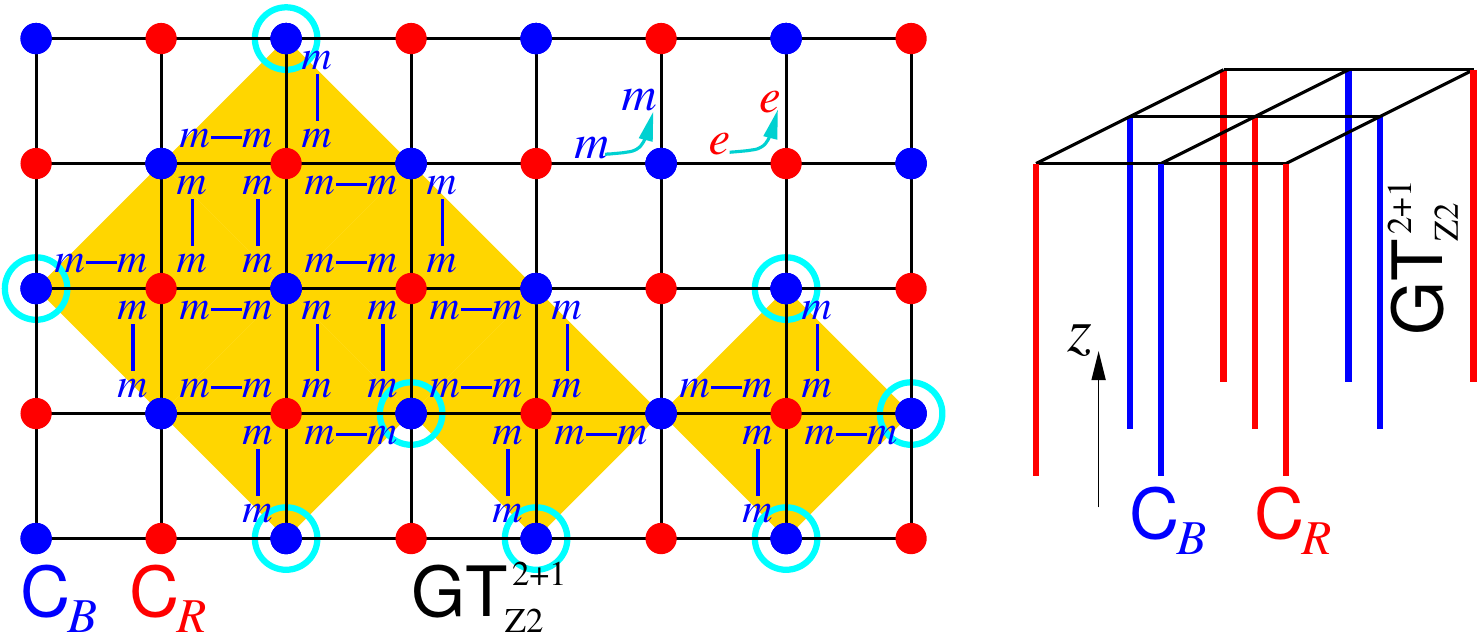} \end{center}
\caption{
A configuration of the cellular topological state.  The unmarked links are in
sector-1.  The links marked by $m-m$ are in sector-3.
Only the vertices in the blue circle cost energy.
}
\label{Z2RBsq}
\end{figure}

\subsection{A cellular topological state on square column lattice} 

In this section we consider a cellular topological state on a lattice formed by
square columns (see Fig. \ref{Z2RBsq}).  The stripes in the $z$-direction are
occupied by 2+1D $\Z_2$ topological order.  The red and blue vertical lines are
two kinds of boundaries, $\sC_R$ and $\sC_B$, of those $\Z_2$ topological orders:
\begin{align}
\Bulk(\sC_R) = \sGT_{\Z_2}^{2+1}  \boxtimes \sGT_{\Z_2}^{2+1}  \boxtimes \sGT_{\Z_2}^{2+1}  \boxtimes \sGT_{\Z_2}^{2+1} ,
\nonumber\\
\Bulk(\sC_B) = \sGT_{\Z_2}^{2+1}  \boxtimes \sGT_{\Z_2}^{2+1}  \boxtimes \sGT_{\Z_2}^{2+1}  \boxtimes \sGT_{\Z_2}^{2+1} .
\end{align}
The two boundaries are characterized by the following condensing topological
excitations (the generators):
\begin{align}
 \sC_R:&\ \  e_1e_2,\ e_1e_3,\ e_1e_4,\ m_1m_2m_3m_4,\ f_1f_2f_3f_4 ,
\nonumber\\
 \sC_B:&\ \  m_1m_2,\ m_1m_3,\ m_1m_4,\ e_1e_2e_3e_4,\ f_1f_2f_3f_4 .
\end{align}

From the condensing particles on the boundaries, we see that the $e$-particle
can move across the $\sC_R$ boundaries, while the $m$-particle can move across
the $\sC_B$ boundaries.  But, the arrangement of the $\sC_R$ and $\sC_B$
boundaries is such that the  long distant motion of the $e$- and $m$-particles
is blocked and they cannot move freely in the $x$-$y$ direction.  This suggests
the cellular topological state to be a non-liquid state.

If all the links are  in sector-1, then we have a minimal energy ground state.
If we change some links to sector-3 (marked by $m-m$ in Fig. \ref{Z2RBsq}), we
will get an excited state.  We may  group the sector-3 links into small
diamonds (see Fig.  \ref{Z2RBsq}).  Only the vertices that touch an odd number
of the diamonds cost a finite energy (see Fig.  \ref{Z2RBsq}), and correspond
to a fracton.  A fracton cannot move by itself in $x$-$y$ directions.  Only a
pair of fractons can move in a certain way in $x$-$y$ directions. But a fracton
can move freely and independently in $z$-direction.  The fractons are created
at the corner of diamond-shaped membrane operators.  Those properties suggests
that the constructed cellular topological state is a type-I fracton state.

\subsection{A cellular topological state on cubic lattice} 

\begin{figure}[t]
\begin{center}
\includegraphics[scale=0.25]{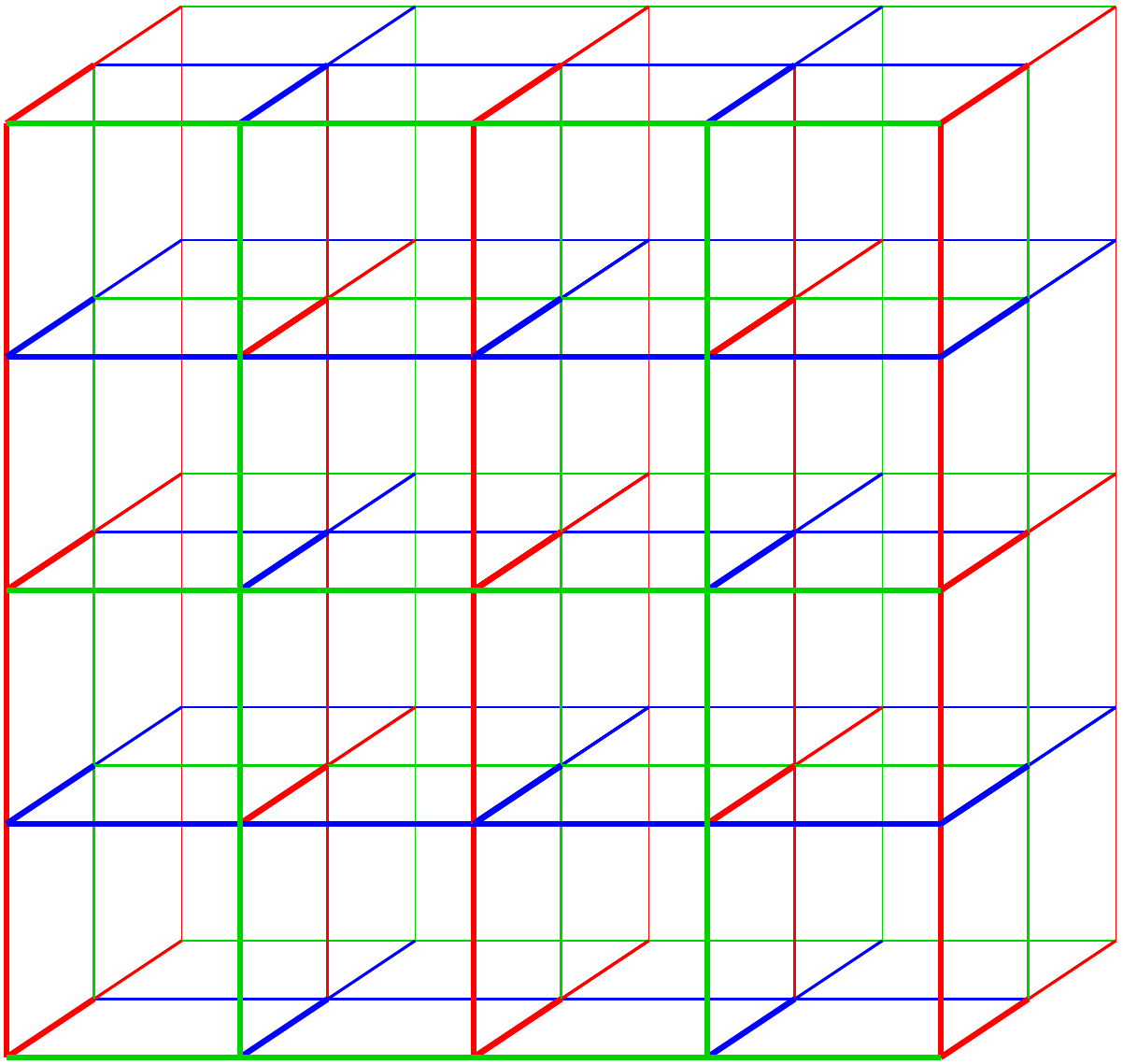} \end{center}
\caption{ A cellular topological state on cubic lattice.  The square faces are
occupied by 2+1D $\Z_2$ topological order, and the red, blue, green lines
correspond to three kinds of 1+1D anomalous topological orders $\sC_R$,
$\sC_B$, and $\sC_G$.  For example, we assign the anomalous topological orders
$\sC_R$ and $\sC_B$ to the links in the $x$-direction in an alternative way.  }
\label{Z2RBGcube}
\end{figure}

Now, we consider a cellular topological state on a cubic lattice columns (see
Fig. \ref{Z2RBGcube}), which is a generalization of the square column model in
the last section.  The square faces of the cubic lattice are occupied by 2+1D
$\Z_2$ topological order.  The red, blue, and green lines are three kinds of
boundaries, $\sC_R$, $\sC_B$, and $\sC_G$ of those $\Z_2$ topological orders:
\begin{align}
\Bulk(\sC_R) = \sGT_{\Z_2}^{2+1}  \boxtimes \sGT_{\Z_2}^{2+1}  \boxtimes \sGT_{\Z_2}^{2+1}  \boxtimes \sGT_{\Z_2}^{2+1} ,
\nonumber\\
\Bulk(\sC_B) = \sGT_{\Z_2}^{2+1}  \boxtimes \sGT_{\Z_2}^{2+1}  \boxtimes \sGT_{\Z_2}^{2+1}  \boxtimes \sGT_{\Z_2}^{2+1} ,
\nonumber\\
\Bulk(\sC_G) = \sGT_{\Z_2}^{2+1}  \boxtimes \sGT_{\Z_2}^{2+1}  \boxtimes \sGT_{\Z_2}^{2+1}  \boxtimes \sGT_{\Z_2}^{2+1} .
\end{align}
The above three boundaries are characterized by the following condensing
topological excitations (the generators):
\begin{align}
 \sC_R:&\ \  e_1e_2,\ e_1e_3,\ e_1e_4,\ m_1m_2m_3m_4,\ f_1f_2f_3f_4 ,
\nonumber\\
 \sC_B:&\ \  m_1m_2,\ m_1m_3,\ m_1m_4,\ e_1e_2e_3e_4,\ f_1f_2f_3f_4 ,
\nonumber\\
 \sC_G:&\ \  f_1f_2,\ f_1f_3,\ f_1f_4,\ e_1e_2e_3e_4,\ m_1m_2m_3m_4 .
\end{align}

From the condensing particles on the boundaries, we see that the $e$-particle
can move across the $\sC_R$ boundaries, the $m$-particle can move across the
$\sC_B$ boundaries,  and the $f$-particle can move across the $\sC_G$
boundaries.  But, the arrangement of the $\sC_R$, $\sC_B$, and $\sC_G$
boundaries is such that the long distant motion of the $e$-, $m$-, and
$f$-particles is blocked and they cannot move freely in the any directions.  In
other words, they are localized in a finite region.  To move further, those
particles must split into more and more particles.  This suggests the cellular
topological state in Fig. \ref{Z2RBGcube} to be a non-liquid state.  In
particular, the structure described in Fig. \ref{Z2RBsq} also appears in the
cubic cellular model, and gives rise to point-like excitations with constrained
motion.

\section{Reverse renormalization and generic construction} 

To systematically understand and to classify a gapped liquid state (such as a
topologically ordered state), we perform wavefunction renormalization by
\emph{removing} the unentangled degrees of
freedom.\cite{VC0466,V0705,AV0804,GLW0809} We hope to obtain a fixed-point wave
function which gives us a classifying understanding of the gapped liquid
states.  We also hope the fixed-point wave function is described by a
topological quantum field theory which does not dependent on the lattice
details.

However, for gapped non-liquid states, due to their intrinsic foliation or
cellular structure, above general approach does not work.  In particular, we
should not expect to have a quantum field theory to describe a non-liquid
state. (But a quantum field theories with explicit layer structure may
work.\cite{SW181201613}) On the other hand, we still hope to obtain some kind
of fixed-point wave functions for non-liquid states, so that we can have a
systematic and classifying understanding of non-liquid states.

Here we like to propose a reverse renormalization approach to obtain the
fixed-point wave functions for non-liquid states.  In such an approach, we
\emph{add} unentangled degrees of freedom to our systems, to separation the
layers in the foliation or cellular structure.  After many steps of
renormalization, we get 3+1D gapped liquid states between layers (\ie within a
cell), such as topologically ordered states or SET/SPT states if we have
symmetry. (In our previous discussions, we have assumed the 3+1D gapped liquid
states to be trivial product states.) On the layers, we have 2+1D anomalous
topological orders, which are the domain walls separating the neighboring 3+1D
topological orders.  The layers join at edges, which correspond to 1+1D
anomalous topological orders. The edges join at vertices, which correspond to
0+1D anomalous topological orders (see Fig.  \ref{cell3d}).

The above reverse renomalization understanding of non-liquid states suggests
the following general construction.  We first decompose the 3d space into cells
(see Fig. \ref{cell3d}).  We assign (possibly different) 3+1D topological
orders to the 3d cells, assign 2+1D anomalous topological orders to the 2d
surfaces, assign 1+1D anomalous topological orders to the 1d edges, and assign
0+1D anomalous topological orders to the 0d vertices.  (Without symmetry, the
0+1D anomalous topological orders are always trivial.\cite{KW1458}) This is a
quite general construction, which may cover all the non-liquid states.
However, some constructions may give rise to ground state degeneracies that can
be lifted by local operators.  We need to include those local operators to lift
the degeneracies and to stabilize the constructed states.  Also, different
constructions may lead to the same gapped non-liquid phase.  Finding the
equivalence relations between different constructions is an every important
issue.

Our construction also works if there are on-site symmetry, by
requiring the (anomalous) topological orders in various dimensions to have the
same symmetry.  In the presence of space group symmetry, we need to choose the
cellular structure to have the space group symmetry.  We also need to choose
(anomalous) topological orders in various dimensions to have the proper
symmetries, as discussed in
\Ref{CLV1301,SH160408151,HH170509243,SQ181011013}.

After posting this paper, the author became aware of a prior unpublished work
(now posted as \Ref{AW200205166}) where a very similar construction, based on
defect network in a 3+1D topological quantum field theory, was proposed.  The
defect planes and defect lines correspond to the (anomalous) 2+1D and 1+1D
topological orders in this paper.  Later, another similar construction
was proposed in \Ref{W200212932}.

I would like to thank Xie Chen and Kevin Slagle to bring the above work to my
attention.  This work is motivated by the presentations in the Annual Meeting
of Simons Collaboration on Ultra-Quantum Matter, where the issue of the fixed
point field theory for fracton phases were discussed.  This research was
partially supported by NSF DMS-1664412.  This work was also partially supported
by the Simons Collaboration on Ultra-Quantum Matter, which is a grant from the
Simons Foundation (651440).

\bibliography{../../bib/all,../../bib/publst,./local}

\end{document}